\documentclass[prl,twocolumn,notitlepage,superscriptaddress]{revtex4-1}
\usepackage[a4paper,top=1.75cm,bottom=1.75cm,left=1.75cm,right=1.75cm]{geometry}
\usepackage{amsmath}
\usepackage{amssymb}
\usepackage{bm}
\usepackage{enumerate}
\usepackage{hyperref}
\usepackage{mathptmx}
\usepackage[utf8]{inputenc}
\usepackage{float}
\usepackage{graphicx}
\DeclareGraphicsExtensions{.eps, .pdf, .jpg, .png}
\usepackage[usenames,dvipsnames]{color}
\graphicspath{{./}}
\usepackage{bookmath}
\usepackage{mygroups}
\usepackage{mycolors}
\setcitestyle{numbers,square}
\begin{document}
\title{Half-Quantum Vortices in Nematic and Chiral Phases of $^3$He}
\author{Robert C. Regan}
\email{robertregan2018@u.northwestern.edu}
\affiliation{Department of Physics \& Northwestern-Fermilab Center for Applied Physics \& Superconducting Technologies \\ Northwestern University, Evanston, Illinois 60208, USA}
\author{Joshua J. Wiman}
\email{jwiman@chalmers.se}
\affiliation{MC2, Chalmers University, Gothenburg, Sweden} 
\author{J. A. Sauls}
\email{sauls@northwestern.edu}
\affiliation{Department of Physics \& Northwestern-Fermilab Center for Applied Physics \& Superconducting Technologies \\ Northwestern University, Evanston, Illinois 60208, USA}
\date{\today}
\begin{abstract}
We report theoretical results for the stability of half-quantum vortices (HQVs) in the superfluid phases of \He\ confined in highly anisotropic Nafen aerogel.
Superfluidity of \He\ confined in Nafen is the realization of a ``nematic superfluid'' with Cooper pairs condensed into a single p-wave orbital aligned along the anisotropy axis of the Nafen aerogel.
In addition to the nematic phase, we predict a second ``chiral'' phase that onsets at a lower transition temperature. This chiral phase spontaneously breaks time-reversal symmetry and is a topological superfluid. Both superfluid phases are equal-spin pairing condensates that host arrays of HQVs as equilibrium states of rotating superfluid \He.
We present results for the structure of HQVs, including magnetic and topological signatures of HQVs in both the nematic and chiral phases of \Hen.
\end{abstract}
\maketitle

\vspace*{-5mm}
\noindent{\it Introduction --}
The discovery of superfluidity in liquid \He\ infused into a low-density, highly anisotropic porous aerogel called Nafen was the realization of a ``nematic superfluid'' with Cooper pairs condensed into a single p-wave orbital aligned along the anisotropy axis of the Nafen aerogel~\cite{dmi15}. This nematic superfluid, or ``polar phase'', is not realized as a stable bulk phase of pure \He, but is predicted to be the stable ground state when confined in 100 nm channels~\cite{wim15,wim18}. In the case of \He\ infused into Nafen aerogel, its structure is well understood in terms of a highly porous random solid of long alumina strands, aligned on average, with typical strand diameter of order 8-9 nm and interstrand distances of order 30-50 nm~\cite{dmi15}. The Nafen structure provides the uniaxial confinement that stabilizes the polar phase of \He~\cite{aoy06,sau13,wim19}.
In lower density Nafen the effect of weaker confinement is to allow Cooper pairs with in-plane orbitals, $\hat{\vp}_{x,y}$, to nucleate, leading to a phase transition from the polar phase to an A-like phase, or chiral phase, with a strong polar distortion, hereafter referred to as the ``polar-distorted chiral phase''.
These newly stabilized phases of \Hen\ are also condensates of spin-triplet Cooper pairs with equal amplitudes for two oppositely aligned spin polarization states, $\ket{\uparrow\uparrow}$ and $\ket{\downarrow\downarrow}$. Thus, they belong to the class of equal-spin pairing condensates that can support ``half-quantum vortices'' (HQVs), topological defects with \emph{one half} the usual quantum of circulation predicted for vortices in a superfluid, i.e. $\nicefrac{1}{2}\left(h/2\,m_3\right)$, where $h$ is Planck's constant and $m_3$ is the mass of the \He\ atom~\cite{vol76}.

Indeed the discovery of HQVs in \He\ infused into Nafen was reported soon after the discovery of the polar phase based on the nuclear magnetic resonance (NMR) signature of pairs of HQVs created by rotating the polar phase of superfluid \Hen~\cite{aut16}. In addition, these authors observe that the NMR signature of the HQV pair persists into the polar-distorted chiral phase of \Hen.
That discovery comes 40 years after HQVs were predicted theoretically as a novel class of topological defects in condensed matter~\cite{vol76}.
The prediction of HQVs, combined with more recent theoretical ideas for developing topological condensed matter as platforms for quantum information processing~\cite{iva01}, led to searches for HQVs in diverse condensed matter systems, from Bose-Einstein condensates of optically trapped spin S=1 $^{23}$Na atoms~\cite{seo15} to spin-triplet superconductors thought to be electronic analogs of superfluid \Hea~\cite{jan11}.

Here we report theoretical predictions for the structure of HQVs, their stability and the pressure-temperature phase diagram for two phases of rotating superfluid \He\ confined in Nafen aerogels which host HQVs.
Our analysis is based on an anisotropic impurity model for Nafen combined with strong-coupling Ginzburg-Landau (GL) theory that quantitatively accounts for the relative stability of the confined equal-spin pairing (ESP) superfluid phases in \Hen\ reported in Ref.~\cite{dmi15}. This strong-coupling GL formalism also accounts for the relative stability of the A and B phases of pure superfluid \He\ over the entire pressure-temperature range~\cite{wim16}, as well as the vortex phase diagram of rotating \Heb~\cite{reg20}.
Using the strong-coupling GL theory, with the addition of the impurity model for the Nafen described below, we predict the pressure-temperature phase diagram for superfluid \Hen. The theoretically predicted phase diagram is in excellent agreement with the experimentally reported phase transitions observed in the ESP phases of \Hen~\cite{dmi15}.

For rotating \Hen\ with $\vOmega\parallel\hat\vz$ we find two distinct vortex phases within the polar and polar-distorted chiral phases. In the polar phase region we find a stable array of pairs of HQVs. The cores of the HQVs are found to be the spin-polarized $\beta$-phase. At lower temperatures HQVs with additional internal structure are found embedded in the polar-distorted chiral phase. This HQV phase is the equilibrium state of rotating \Hen\ at temperatures below the polar to chiral phase transition, which is characterized by an anisotropic chiral order parameter and spontaneous supercurrents flowing \emph{along the axis of the HQVs} and confined near their cores. Observation of these effects would provide key signatures for the identification of HQVs in both the polar and chiral phases of \Hen.

\medskip\noindent{\it Order parameter --}
The GL free energy is a functional of the order parameter, the pairing self energy for the condensate amplitude of Cooper pairs, $\Delta_{\sigma\sigma'}(\vp)=g\,\langle \psi_{\sigma}(\vp)\psi_{\sigma'}(-\vp)\rangle$ where $g>0$ is the pairing interaction in the p-wave, spin-triplet channel. Thus, $\hat\Delta(\vp)=\left(i\sigma_{\alpha}\sigma_y\right)\cdot d_{\alpha}(\vp)$ is a $2\times 2$ matrix in spin space,
\begin{eqnarray}
\begin{pmatrix} 
\Delta_{\uparrow\uparrow}	&	\Delta_{\uparrow\downarrow}
\\
\Delta_{\uparrow\downarrow}	&	\Delta_{\downarrow\downarrow}
\end{pmatrix}
=
\begin{pmatrix} 
-d_x(\vp) + i d_y(\vp)		&	d_z(\vp)
\\
d_z(\vp)			&	d_x(\vp) + i d_y(\vp)
\end{pmatrix}
\,,
\end{eqnarray}
where $d_{\alpha}(\vp)$ for $\alpha=\{x,y,z\}$ transforms as a vector under spin rotations, and can be expanded in the vector basis of $l=1$ spherical harmonics, $d_{\alpha}(\vp)=\sum_{i=x,y,z}A_{\alpha\,i}\,\hat{\vp}_i$. Thus, a general spin-triplet, p-wave condensate is described by a $3\times3$ matrix of complex amplitudes, $A_{\alpha i}$, that transform as the vector representation of $\point{SO(3)}{S}$ with respect to the spin index $\alpha$, and as the vector representation of $\point{SO(3)}{L}$ with respect to the orbital momentum index $i$. This 18 dimensional order parameter space allows for a wide variety of topologically stable defects in superfluid \He~\cite{sal87,vollhardt90}.

A special class of {\it equal-spin pairing} (ESP) states are those for which $\Delta_{\uparrow\downarrow}=0$ and $\vert\Delta_{\uparrow\uparrow}\vert=\vert\Delta_{\downarrow\downarrow}\vert$ for all $\vp$ for a fixed direction $\hat\vd$ in spin space.
For \He\ confined in low density Nafen, there are two bulk ESP phases: 
(1) the polar phase described by an order parameter of the form,
$A^{\text{P}}_{\alpha i}=\DeltaP\,\hat\vd_{\alpha}\,\hat\vn_{i}$, where $\hat\vn$ is the direction of the 
orbital pair wave function,
and  
(2) the chiral phase described by an order parameter of the form,
$A^{\text{C}}_{\alpha i}=\DeltaC\,\hat\vd_{\alpha}\,\left(\hat\vn_{i} \pm i \epsilon\hat\vm_{i}\right)$, 
where a second p-wave orbital develops with $\hat\vm\perp\hat\vn$ and out of phase by $\pm\pi/2$. This type of \emph{in-plane} chiral phase was also found to be a stable equilibrium phase in $100\,\mbox{nm}$ cylindrical pores \cite{wim15}. 

The orbital axis $\hat\vn$ of the polar phase is locked along the anisotropy axis of Nafen, i.e. $\hat\vn = \hat\vz$. 
The direction of the spin quantization axis, $\hat\vd$, is weakly coupled to the orbital state via the nuclear dipolar energy which aligns $\hat\vd\perp\hat\vn$, i.e. $\hat\vd=\cos\alpha\hat\vx+\sin\alpha\hat\vy$. The angle $\alpha\in\{0,2\pi\}$ is thus a degeneracy variable for the ESP phase. 
The other degeneracy variable is the global phase, $\vartheta\in\{0,2\pi\}$, of the amplitude, $\Delta=|\Delta|e^{i\vartheta}$ for either the polar or chiral phase.

\medskip\noindent{\it Half-Quantum Vortices --}
Quantized vortices with global phase winding, $\Delta\vartheta=2\pi$, correspond to vortices with the standard quantum of circulation, $\kappa=\oint\,\vv_s\cdot d\vl = h/2m_3$, where $\vv_s = \nicefrac{\hbar}{2m_3}\grad\vartheta$ is the superfluid velocity field. 
For ESP phases, vortices with half the standard quantum of circulation, i.e. global phase winding $\Delta\vartheta=\pm\pi$, are possible. These HQVs are topologically stable line defects in which the sign change resulting from the $\pi$ phase winding, $\Delta\xrightarrow[]{\cC}-\Delta$ is compensated by a sign change in the direction of the spin quantization axis, $\hat\vd\xrightarrow[]{\cC}-\hat\vd$, upon traversing a closed circuit $\cC$. The far-field structure of the HQV is particularly clear in the ESP basis~\cite{sau16},
\begin{equation}
\Delta_{\uparrow\uparrow}    = |\Delta|e^{i(\vartheta+\alpha)}\,\cY(\vp) 
\,,\quad
\Delta_{\downarrow\downarrow} = |\Delta|e^{i(\vartheta-\alpha)}\,\cY(\vp) 
\,,
\label{eq-Delta_ESP}
\end{equation}
where $\cY(\vp)$ is the orbital order parameter, e.g. $\cY(\vp)=\hat\vn\cdot\hat\vp$ for the polar state, and $\cY(\vp)=\left(\hat\vn+i\varepsilon\hat\vm\right)\cdot\hat\vp$ for the chiral phase.
The spin-polarized amplitudes depend on the phase variables, 
$\vartheta_{+}=\vartheta+\alpha$ and 
$\vartheta_{-}=\vartheta-\alpha$.
Thus, there are two distinct HQVs with $\Delta\vartheta=\pi$ corresponding to 
$\Delta\alpha=+\pi$ or $\Delta\alpha=-\pi$;  equivalently 
$\Delta\vartheta_{+} = 2\pi$ and  
$\Delta\vartheta_{-} = 0$, or
$\Delta\vartheta_{+} = 0$ and  
$\Delta\vartheta_{-} = 2\pi$, respectively. 
Thus, an HQV is a $2\pi$ phase vortex in only one of the two ESP condensates, which accounts for mass circulation of half the normal value. It is also clear that the two types of HQVs correspond to spin current vortices with opposite spin polarizations. Equations \ref{eq-Delta_ESP} correspond to the far-field asymptotic forms of the order parameter of the HQVs. 
The general form for the HQV order parameter is expressed in terms of the full matrix order parameter, $A_{\alpha i}(\vr)$. We first discuss the stabilization of the polar and polar-distorted chiral phases of superfluid \He\ infused into Nafen aerogel.

\medskip\noindent{\it Impurity Model for \He\ in Nafen -- }
Ginzburg-Landau theory has been formulated to calculate the order parameter and thermodynamic properties of {\it inhomogeneous} phases of superfluid \He~\cite{wim15,wim16,wim18,reg20}. Here we develop the strong-coupling formulation of GL theory for superfluid \He\ infused into Nafen aerogels.

The Ginzburg-Landau free energy functional is expressed in terms of linearly independent invariants constructed from $A_{\alpha i}$, $A_{\alpha i}^{*}$ and their gradients, $\grad_j A_{\alpha i}$ and $\grad_j A_{\alpha i}^{*}$. In particular, the GL functional can be expressed in terms of free energy densities,
\begin{equation}\label{eq-GL_functional}
\hspace*{-2mm}
\cF[A]
\ns=\ns
\int_{V}\ns d^3r
\left\{
f_{\mathrm{b}}[A] 
+ 
f_{\mathrm{Z}}[A]
+ 
f_{\mathrm{d}}[A]
+ 
f_{\text{$\grad$}}[A] 
+
f_{\mathrm{imp}}[A]
\right\}
\,,
\end{equation}
where the bulk free energy density, $f_{\mathrm{b}}$, is given by one second-order invariant and five fourth-order invariants, the nuclear Zeeman energy, $f_{\text{Z}}$, and nuclear dipole-dipole energy, $f_{\mathrm{d}}$, which are also second-order in $A$, and the gradient energy, $f_{\text{$\grad$}}$, which is second-order in gradients of $A$. These terms, and the pressure-dependent material coefficients that define the strong-coupling GL functional, are discussed in detail in Ref.~\cite{reg20} and summarized in the Appendix.
The last term in Eq.~\ref{eq-GL_functional}, $f_{\mathrm{imp}}$, is the free energy density associated with pair breaking by the Nafen strands. Impurities in a p-wave superfluid are pair breaking. Elastic scattering of quasiparticles comprising Cooper pairs leads to suppression of the order parameter over a region of order the coherence length, $\xi$, and a loss in condensate energy that depends on the density of impurities and the quasiparticle-impurity cross-section.
Nafen is a highly porous anisotropic material comprised of long strands of crystalline Al$_2$O$_3$, c.f. Fig.~1 of Ref.~\cite{asa15}. Nafen-90, with density, $\rho=90\,\mbox{mg/cm}^3$, has a mean inter-strand distance of $L_s=47.8\,\mbox{nm}$, strand lengths of order millimeters and mean strand radius of $r_s=4.0\,\mbox{nm}$~\cite{asa15}.

For our analysis of the effects of Nafen on liquid \He\ infused into the Nafen structure, we model Nafen as an array of non-magnetic line impurities of local areal density $n_s(\vr)$. The effects of the array of anisotropic impurities on the superfluid phases of \He\ is described to leading order in $A$ by the pair-breaking free energy density, 
\begin{equation}\label{eq-GL_impurity}
f_{\mathrm{imp}} = A_{\alpha i}\,\mrfI_{ij}(\vr)\,A_{\alpha j}^{*}
\,,
\end{equation}
where $\mrfI_{ij}(\vr)$ is a uniaxial tensor under orbital space rotations that depends on the local areal strand density, $n_s(\vr)$. In general the nematic axis varies in space. Here we treat the Nafen strands as globally aligned along an axis, $\hat\vz$, and neglect fluctuations in the orientation of the nematic axis~\footnote{The effects of random anisotropy in anisotropic aerogels on the phases of superfluid \He\ is discussed by several authors. See Ref.~\cite{sau13} and references therein.}. Thus, for uniaxial nematic aerogels the impurity free energy density is defined by 
\begin{eqnarray}
\mrfI_{ij}(\vr)
&=&
\onethird\,N_f\,\xi_0\,n_s(\vr)\,\sigma_{ij}
\,,
\label{eq-Nafen_impurity_tensor}
\\
\mbox{where}\quad 
\sigma_{ij}
&=&
\left[
\sigma_{\perp}\left(\delta_{ij}-\hat{z}_i\hat{z}_j\right)
+
\sigma_{\parallel}\,\hat{z}_i\hat{z}_j
\right]
\,,
\label{eq-Nafen_impurity_cross-sections}
\end{eqnarray}
is a uniaxial tensor that parametrizes the anisotropic scattering of quasiparticles by Nafen strands with areal density, $n_s(\vr)$. The coefficients $\sigma_{\perp}$ and $\sigma_{\parallel}$ determine the cross-radii for quasiparticle scattering normal to the impurity strands and at grazing incidence along the strands, respectively. This form for $I_{ij}$ is consistent with the pair-breaking energy resulting from scattering by impurities embedded in superfluid \He\ in Refs.~\cite{rai77,sau13}.
The prefactor, $\tinyonethird\,N_f\,\xi_0$, is the scale set by the coefficient of the second-order term for the bulk condensation energy, i.e. $f_{\mathrm{b}}=\alpha(T)\Tr{AA^{\dag}}$ with $\alpha(T)=\tinyonethird\,N_f\,\ln(T/T_{c})$, and the coherence length $\xi_0\equiv \hbar v_f/2\pi\kb T_c$ that enters the GL material parameters for the gradient energy density, $f_{\text{$\grad$}}$ (see Eqs. 4 and 9 of Ref.~\cite{reg20}). 
Note that $T_c$ is the superfluid transition temperature of pure, bulk \He, $N_f=\tinythreefourths\,n/E_f$ is the normal state density of states at the Fermi level, $E_f=\tinyonehalf p_f v_f$ is the Fermi energy, $p_f$ is the Fermi momentum, $v_f=p_f/m^*$ is the Fermi velocity, $m^*$ is the effective mass of the \He\ quasiparticles, and $n=p_f^3/3\pi^2\hbar^3$ is the density of liquid \He. 
All of these Fermi liquid properties, as well as the stiffness coefficients, $K_{1,2,3}$, defining the GL gradient energy, $f_{\text{$\grad$}}$ (Eqs.~4 and 9 of Ref.~\cite{reg20}) and the strong-coupling $\beta$ parameters (Eqs.~2,7,8 and 12 of Ref.~\cite{reg20}) defining the bulk free energy density, $f_{\mathrm{b}}$, depend on pressure and are given in, or can be obtained from, Table II of Ref.~\cite{reg20}.

The second-order impurity contribution given by Eq.~\ref{eq-GL_impurity} is essential, not only because it breaks rotation symmetry, but also because it competes with $\alpha(T,p)\,A_{\mu i}\,A_{\mu i}^*$, where $\alpha(T,p)$ is vanishingly small in the GL limit, i.e. near the bulk $T_{c}$.
Uniaxial symmetry imposed on $^3$He by Nafen allows for additional anisotropic impurity corrections to the the gradient energy density, $f_{\mbox{\tiny $\nabla$}}$, as well as the fourth-order bulk free energy density, $f_{\mbox{\tiny b}}$.
However, the gradient coefficients, e.g. $K_1\sim N_f\xi_0^2$, are finite near $T_{c}$. Thus, Nafen with mean impurity density, $\bar{n}_s=1/L_s^2$, generates perturbative corrections to the gradient energies of relative order $f_{\text{$\nabla$}}^{\text{imp}}/f_{\text{$\nabla$}}=\sigma_{||,\perp}\xi_0/L_s^2$, which varies from $0.06$ at high pressure to $0.20$ at $p=0$ bar. 
Thus, the transition line $T_{c_2}(p)$ shown in Fig.~\ref{fig-Phase_Diagram}, as well as the spatial variations of the order parameter, are well described by retaining the leading order impurity term in the GL functional, Eq.~\ref{eq-GL_impurity}. The uniaxial corrections to the gradient terms and fourth order terms may be related to the small deviations from experiment in the calculated transition line $T_{c_2}(p)$ at the lowest pressures, as shown in Fig.~\ref{fig-Phase_Diagram}, but the analysis required to check this conjecture is outside the scope of this paper.

\medskip\noindent{\it Phase Diagram of \Hen\, --}
Nematic impurities break the orbital rotation symmetry of pure normal \He. As a result they split the three-dimensional p-wave representation of $\orbital$ into a one-dimensional ($\hat{\vp}_z$) orbital representation for Cooper pairs aligned with the nematic impurities, and a two-dimensional representation ($\hat{\vp}_x,\hat{\vp}_y$) for Cooper pairs with orbital wave functions normal to the array of nematic impurities.
Pair breaking leads to suppression of the orbital components of the p-wave Cooper pairs, and a corresponding loss in condensation energy. In particular for $\sigma_{\perp} > \sigma_{\parallel} \ge 0$ there is stronger suppression of the orbital components in the plane perpendicular to the strands, i.e. $\hat{\vp}_{x,y}$, compared to that for Cooper pairs in the $\hat{\vp}_z$ orbital state.
In this limit we expect the onset of superfluidity into the \emph{polar phase} with p-wave orbital $\hat{\vp}_z$.

The areal density for impurities is given by $n_s(\vr) = \sum_{i=1}^{N_s}\delta^{(2)}(\vr - \vr_i)$, where $\vr_i$ is the position in the two-dimensional plane of the $i^{\text{th}}$ line impurity and $N_s$ is the total number of Nafen strands. The local density of a single line impurity is reasonably represented by the two-dimensional delta function when the geometric radius of the Nafen strand is small compared to the coherence length, i.e. $r_{s}\ll\xi_0$, which is the case for \He\ infused into Nafen aerogel~\footnote{See Eq.~9.14 and related text in Ref.~\onlinecite{ser83}.}.
\begin{figure}[t]
\centering
\includegraphics[width=\columnwidth]{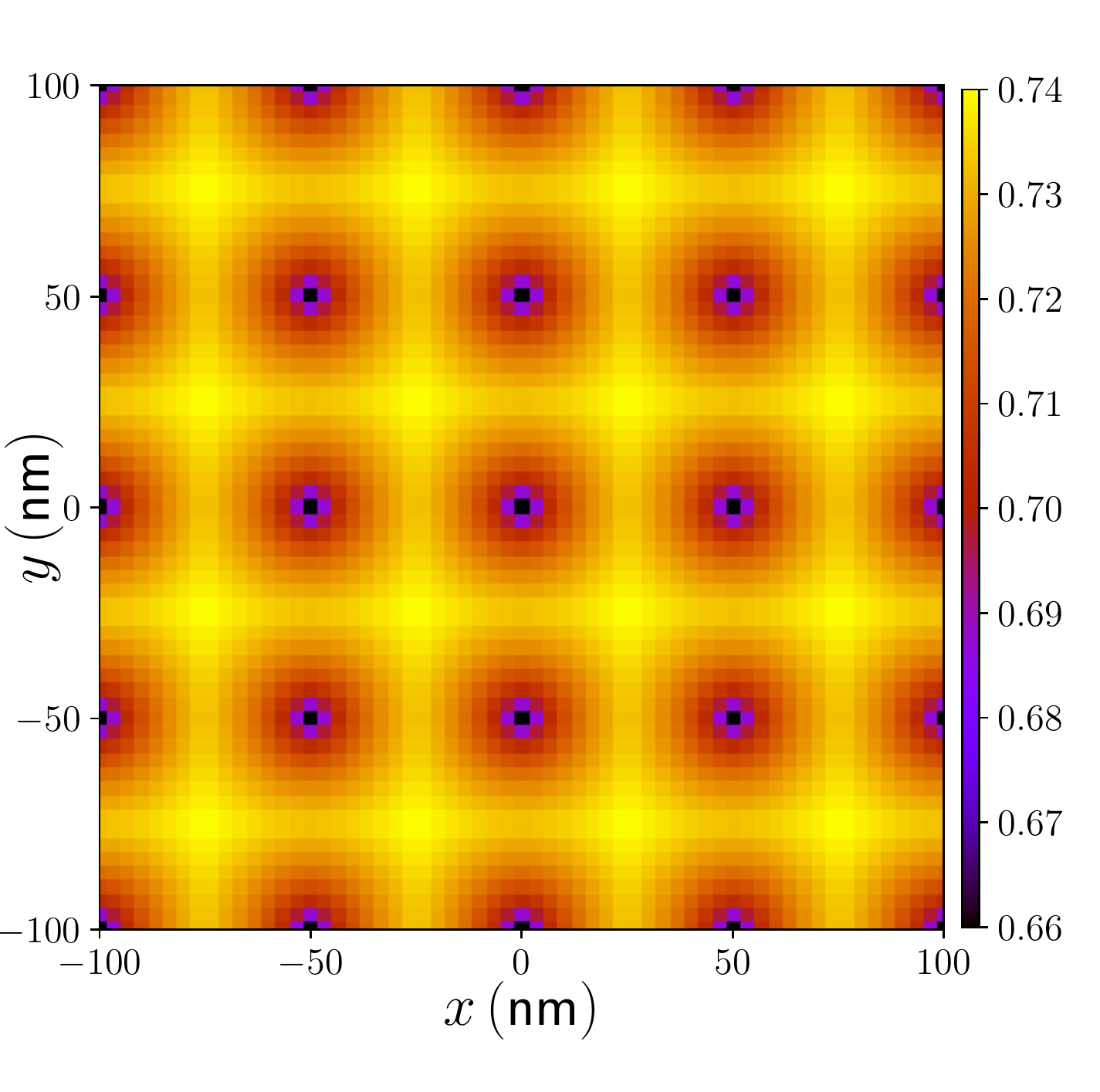}
\caption{Polar phase order parameter amplitude, $\Delta^{\text{P}}(\vr)$, embedded in a square lattice impurity model for Nafen-90. The amplitude is shown for $T=0.95 T_c$ and $p=15$ bar, and is scaled in units of the bulk polar amplitude, $\Delta_{\text{P}}=\sqrt{|\alpha(T)|/2\beta_{12345}}$. The parameters for the lattice model for Nafen-90 are described in the text. 
\label{fig-Nafen90_Lattice_Model}
}
\end{figure}

We investigate the pair-breaking effects of Nafen aerogel on superfluid \He\ by embedding \He\ in a square lattice of line impurities with mean areal density $\bar{n}_{s} = 1/L_{s}^2$ where $L_s$ is the average interstrand distance of Nafen, c.f. Figure~\ref{fig-Nafen90_Lattice_Model}.
The second-order transition from the normal state to the polar state is obtained from linear instability analysis as described in the Appendix. The transition temperature is generally suppressed and given by
\be
T_{c_1} = T_c \left[1-\left(\bar{n}_s\xi_0^2 + \beta\right)\frac{\sigma_{\parallel}}{\xi_0}\right]
\,,
\label{eq-Tc1}
\ee
to leading order in $\sigma_{\parallel}/\xi_0$, where $\bar{n}_s\xi_0^2$ is the dominant contribution from the $Q=0$ mode, while $\beta\approx 0.19\ln(0.10\,L_s/r_s)$ is a correction from the $Q\ne 0$ modes (see Appendix).
Note that for $\sigma_{\parallel} = 0$, corresponding to specular scattering along the strands, i.e. $\vp_z'=\vp_z$, there is no suppression of the superfluid transition relative to that of pure \He; scattering from the strands leads only to suppression of pairing into the in-plane $\vp_x,\vp_y$ orbitals~\cite{fom18}.

The experimental data (red squares in Fig.~\ref{fig-Phase_Diagram}) for the normal to polar transition in Nafen-90 shows that $T_{c_1}<T_c$ implying non-specular scattering along the strands. The theoretical result (black line) for $T_{c_1}$ from Eq.~\ref{eq-Tc1} is shown in Fig.~\ref{fig-Phase_Diagram} for a pressure independent scattering cross section of $\sigma_{\parallel}=2.565\,\mbox{nm}$. We also used the experimentally reported areal density, or strand spacing $L_s=47.8\,\mbox{nm}$, and strand radius $r_s=4.0\,\mbox{nm}$. 
Note that the pressure dependence of $T_{c_1}$ arises from the pressure dependence of $T_c(p)$ for pure \He\ and the corresponding coherence length, $\xi_0(p)=\hbar v_f/2\pi\,\kb\,T_c$, which were taken from Table II of Ref.~\cite{reg20}.

\begin{figure}[t]
\centering
\includegraphics[width=\columnwidth]{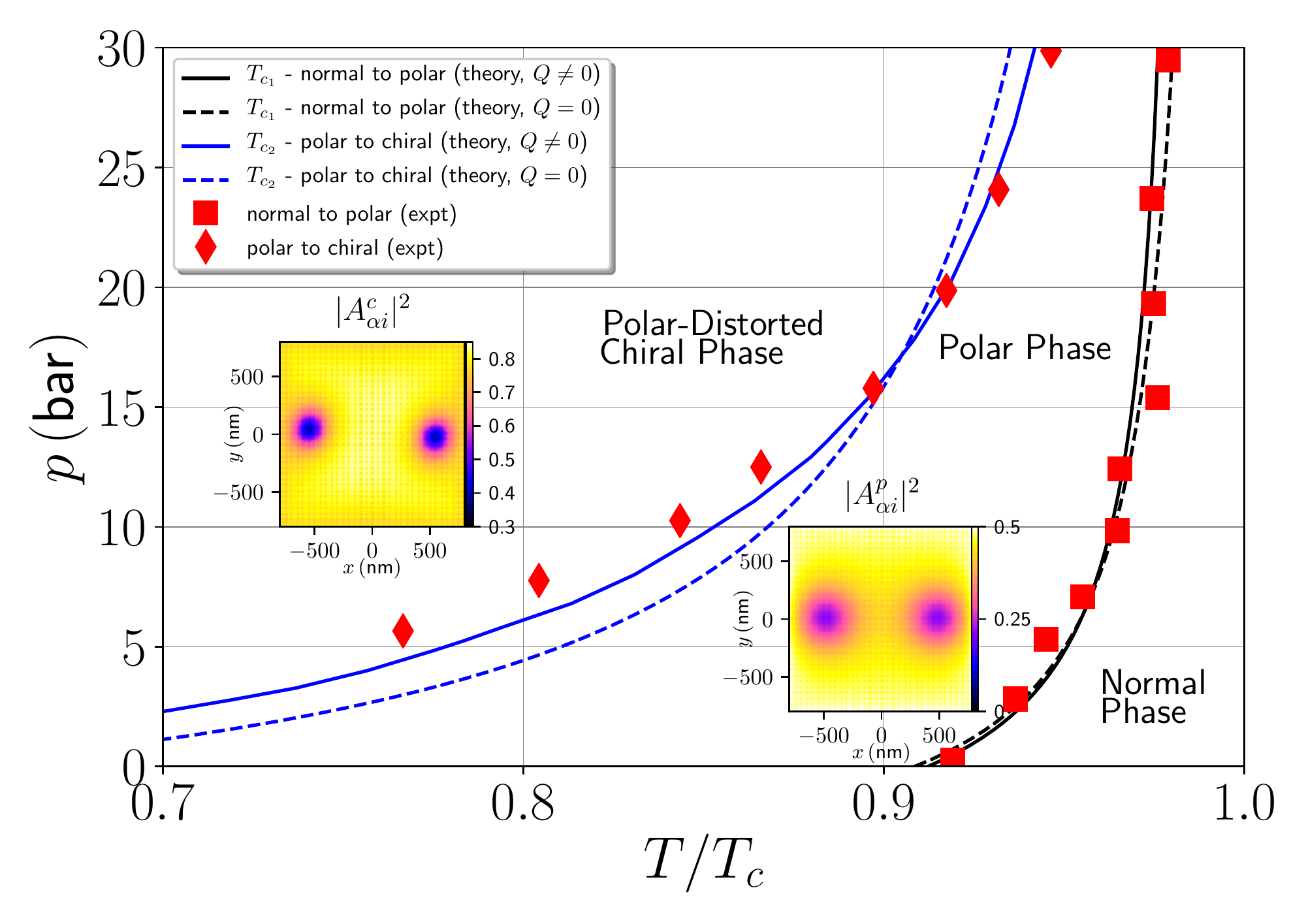}
\caption{
Pressure-temperature phase diagram for \Hen. Experimental data for transitions from the normal to polar phase (red squares) and for the polar to chiral phase (red diamonds) for \He\ in Nafen-90 are reproduced from Ref.~\onlinecite{dmi15}. The solid black curve is $T_{c_1}$ calculated from Eq.~\ref{eq-Tc1} for $\sigma_{||}=2.565\,\mbox{nm}$. The solid blue curve is the transition line $T_{c_2}(p)$ for $\sigma_{\perp}=10.16\,\mbox{nm}$ obtained by numerical minimization of the strong-coupling GL free energy functional. 
The dashed black curve is the best fit of $T_{c_1}$ neglecting the $Q\ne 0$ modes with $\sigma_{||}=2.707\,\mbox{nm}$, while
the dashed blue curve is the result for $T_{c_2}(p)$ based on the $Q=0$ mode given by Eq.~\ref{eq-Tc2} in the Appendix with $\sigma_{||}=2.565\,\mbox{nm}$ and $\sigma_{\perp}=8.021\,\mbox{nm}$. Insets: superfluid density plots for the HQV pairs in the polar and chiral phases.
}
\label{fig-Phase_Diagram}
\vspace*{-5mm}
\end{figure}

For $T_{c_2} < T < T_{c_1}$ an inhomogeneous polar phase is the equilibrium phase with order parameter, $A_{\alpha i} = \Delta^{\text{P}}(\vr)\,\hat{d}_{\alpha}\hat{z}_i$. The inhomogeneity of the polar phase order parameter, induced by pair-breaking from the Nafen impurities, is shown in Fig.~\ref{fig-Nafen90_Lattice_Model}.
For Nafen-90 we find a second transition at $T_{c_2}(p)$, calculated and shown in Fig.~\ref{fig-Phase_Diagram} for in-plane scattering cross section $\sigma_{\perp}=10.16\,\mbox{nm}=3.961\times\sigma_{||}$, at which the Cooper pairs generate $\hat{\vp}_{x,y}$ orbitals in the presence of a dominant polar amplitude~\footnote{Measurements of spin diffusion of $^3$He quasiparticles in Nafen reported in Ref.~\cite{dmi15b} give a ratio of mean free paths for heat transport along ($\parallel$) or perpendicular ($\perp$) to the Nafen strands of $l_{\parallel}/l_{\perp}=960\,\mbox{nm}/290\,\mbox{nm}=3.31$. The mean free paths for uniaxial Nafen are inversely proportional to the corresponding cross sections for quasiparticle scattering off a \emph{homogeneous density} of line impurities, i.e. $l_{\parallel}/l_{\perp}=\sigma_{\perp}/\sigma_{\parallel}$. Our analysis of the phase diagram for an array of line impurities yields the ratio $\sigma_{\perp}/\sigma_{\parallel}=3.96$, in rough agreement with that inferred from spin diffusion. The comparison is better if we use the value of $\sigma_\perp$ obtained for a {homogeneous aerogel density}, $\sigma_{\perp}/\sigma_{\parallel}=8.021\,\mbox{nm}/2.565\,\mbox{nm}=3.13$.}. An approximate analytic result for $T_{c_2}(p)$ based on the dominant $Q=0$ mode of the linear instability equation is shown for comparison and discussed in the Appendix~\footnote{See also Ref.~\cite{min14}}.
We find an order parameter that is an ESP state with $A_{\alpha i}(\vr)= \hat{d}_{\alpha}\,(\Delta^{\text{P}}(\vr)\hat{z}_i\,\pm i\,\Delta^{\text{$\perp$}}(\vr)\hat{m}_i)$, with $\hat\vm\perp\hat\vz$. This phase breaks time-reversal symmetry with the chiral axis \emph{in-plane}, $\hat\vl = \hat\vz\times\hat\vm$. 
The strong-coupling GL theory and the impurity lattice model for Nafen-90 accurately account for the relative stability of the superfluid phases of \Hen~\footnote{Based on NMR frequency shift the polar-distorted chiral phase is a Larkin-Imry-Ma (LIM) phase in which random transverse fluctuations of the Nafen anisotropy axis lead to finite range orientational correlations of the  transverse orbital state, i.e. $\langle\hat\vm(\vr)\hat\vm(\mathbf{0})\rangle=e^{-r/\xi_{\mbox{\tiny LIM}}}$. The LIM correlation length is of order $\xi_{\mbox{\tiny LIM}}\approx 1\,\mu\mbox{m}$~\cite{dmi15}, and the characteristic wavenumber, $Q_{\mbox{\tiny LIM}}=2\pi/\xi_{\mbox{\tiny LIM}}\ll Q_1=2\pi/L_s$, where $L_s\approx 50\,\mbox{nm}$ is the mean spacing between Nafen impurities.
Thus, the gradient energy associated with LIM fluctuations, $f_{\mbox{\tiny LIM}}\propto K_1 Q_{\mbox{\tiny LIM}}^2$, which in principle contributes to suppression of $T_{c_2}$, is dominated by the gradient energy associated depairing by the lattice of Nafen impurities, $f_{\mbox{\tiny $\nabla$}}\propto K_1 Q_1^2$. Thus, the phase transition line, $T_{c_2}(p)$ shown in Fig.~\ref{fig-Phase_Diagram} is accurately calculated neglecting the LIM fluctuations.}. Strong-coupling corrections to the GL functional are essential for the stability of the polar-distorted chiral phase. Weak-coupling theory predicts a polar to polar-distorted B-phase and no polar-distorted chiral phase~\cite{fom13,tan20}.
Within strong-coupling GL theory, with strong uniaxial anisotropy, the polar to polar-distorted chiral transition persists over the full pressure range for strongly anisotropic Nafen-90. 
Strong-coupling effects weaken at pressures below the bulk triple point, but are still present down to $p=0$ bar.
The polar-distorted B phase is stabilized at lower temperatures for uniaxial anisotropy corresponding to Nafen-90, but we find that the polar-distorted chiral phase always appears between the polar phase and the polar-distorted B phase, if the latter is stable. See Fig.~4.2 and the related discussion in Ref.~\cite{wim19}.

Note that the nuclear dipolar energy (Eq.~5 in Ref.~\cite{reg20}) is minimized for $\hat\vd\perp\hat\vz$ for both the polar phase and the polar-distorted chiral phase, as is the Zeeman energy for fields parallel to the nematic axis, $\vH\parallel\hat\vz$. This is the configuration suitable for topologically stable half-quantum vortices.

\medskip\noindent{\it Half-Quantum Vortex Pairs in the Polar Phase --}
In a rotating cryostat superfluid \He\ co-rotates with the confining cell by nucleating an array of quantized vortices with mean areal density given by the Feynman-Onsager relation, $n_v=2\Omega/\kappa_3$, where $\Omega$ is the angular speed of rotation and $\kappa_3=h/2m_3\simeq 0.67\times 10^{-3}\,\mbox{cm}^2/\mbox{sec}$ is the quantum of circulation for singly quantized vortices. For an array of half-quantum vortices the circulation quantum is $\kappa_3/2$ and thus co-rotation requires twice the areal density of HQVs.
In a cylindrical cell of cross-sectional area 1 cm$^2$ rotating at $1\,\mbox{rev}/\mbox{sec}$ co-rotation requires approximately $1.9\times 10^4$ singly quantized vortices with mean spacing $d_v\approx 70\,\mu\mbox{m}$. By comparison the radial extent of the vortex core is of order $2\xi_0\approx 0.16\,\mu\mbox{m}$. This separation of length scales allows us to introduce a computational cell that is large compared to vortex-core structures in order to determine the relative stability of different vortex states for fixed areal density per circulation quantum.
In particular, for fixed rotation speed we can compare the total energy of an array of singly quantized vortices with an array of HQVs.

There are two distinct HQVs, as described in Eq.~\ref{eq-Delta_ESP}, in which one spin component hosts a $2\pi$ phase vortex while the time-reversed spin component is vortex free. The matrix representation of the order parameter corresponding to the HQV with $2\pi$ phase winding of the $\Delta_{\uparrow\uparrow}$ component is 
$A^{+}_{\alpha i}(\vr) = \nicefrac{1}{2}
\left[
\Delta_{+}(\vr)\,e^{i\phi}\,(\hat{x}+i\hat{y})_{\alpha}\,\hat{z}_i
+
\Delta_{-}(\vr)\,(\hat{x}-i\hat{y})_{\alpha}\,\hat{z}_i
\right]$, 
while that for the HQV with $2\pi$ phase winding in $\Delta_{\downarrow\downarrow}$ is,
$A^{-}_{\alpha i}(\vr) = \nicefrac{1}{2}
\left[
\Delta_{+}(\vr)\,(\hat{x}+i\hat{y})_{\alpha}\,\hat{z}_i
+
\Delta_{-}(\vr)\,e^{i\phi}\,(\hat{x}-i\hat{y})_{\alpha}\,\hat{z}_i
\right]$.
Superfluid rotation via HQVs involves pairs of these spatially separated HQVs, 
\begin{equation}\label{eq-PolarHQV_IC}
\hspace*{-2.5mm}
A_{\alpha i}(\vr) 
\ns=\ns
\frac{1}{2}
\left[
\Delta_{+}(\vr)e^{i\phi_{+0}}(\hat{x}\ns+\ns i\hat{y})_{\alpha}
\ns+\ns
\Delta_{-}(\vr)e^{i\phi_{-0}}(\hat{x}\ns-\ns i\hat{y})_{\alpha}
\right]
\hat{z}_i,
\end{equation}
with $\phi_{\pm 0}=\arctan(y/(x\mp x_0/2))$ where $x_0$ is the distance between the phase singularities of the HQV pair. Note that $\Delta_{\pm}(\vr)$ have zeroes at $\pm x_0/2$, and that for $x_0=0$ the pair of HQVs collapse to a singly quantized vortex (SQV) in the polar phase with spatially uniform $\hat\vd = \hat\vx$.

\begin{figure} 
\centering
\includegraphics[width=\columnwidth]{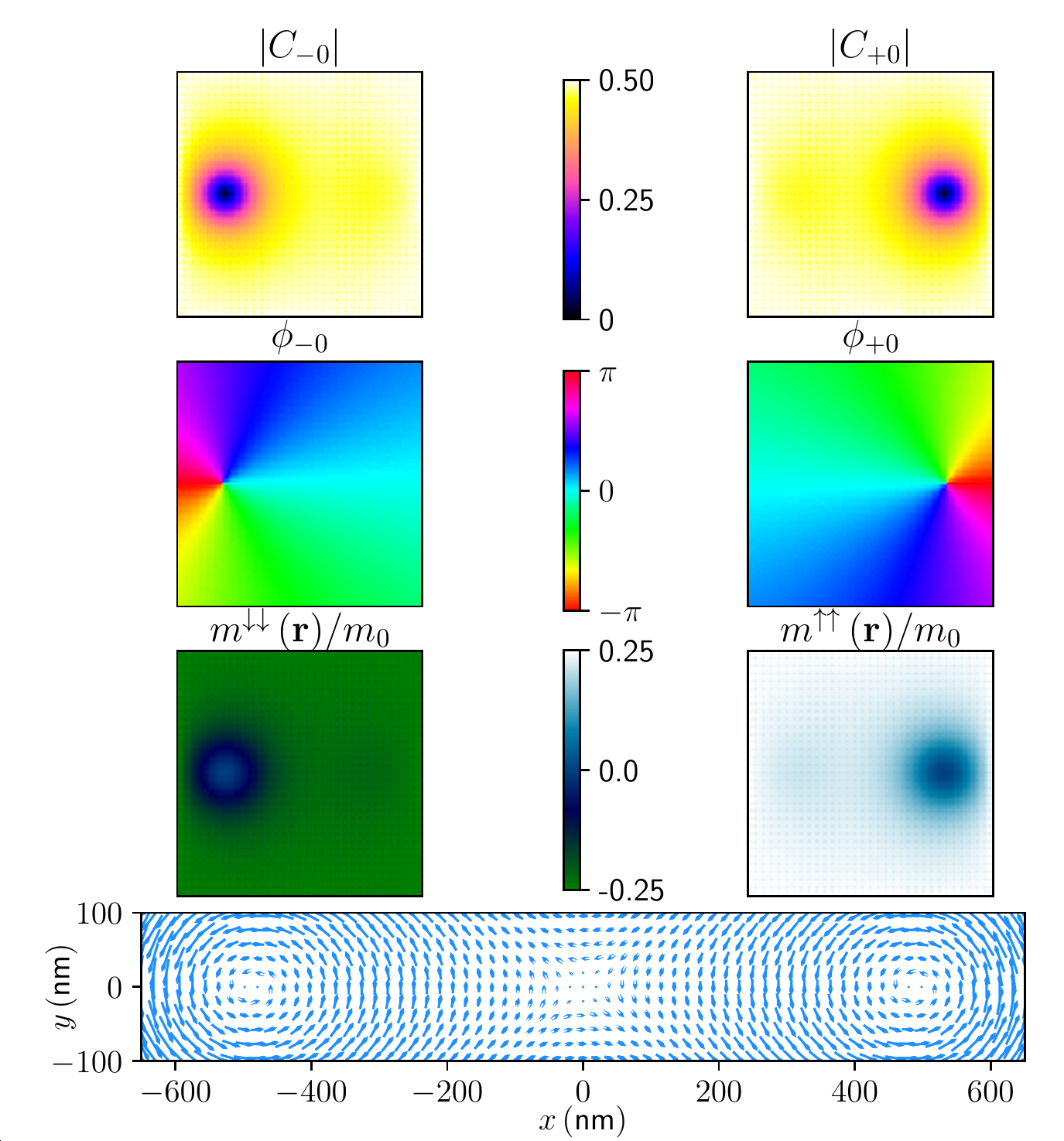}
\caption{Upper panel: Amplitudes for the two components, $|C_{\pm 0}|$, of the HQV pair at $p=15$bar and $T=0.95T_c$ calculated on a square grid with linear dimension $1600\,\mbox{nm}\approx 15\xi(T)$.
The corresponding phase plots showing the $2\pi$ vortices for each component of the HQV pair are shown in the second panel.
Third panel: The magnetization densities of the HQV pair in the ESP basis. 
Bottom panel: Vector plot showing the broken axial symmetry in the mass current density for the HQV pair.
}
\label{fig-HQV_Pair-Polar}
\end{figure}

We calculated the equilibrium and metastable vortex solutions for corresponding pairs of HQVs, as well as SQVs, per unit cell by numerical minimization of the strong-coupling GL free energy functional for vortex phases in both the polar and chiral regions of the phase diagram shown in Fig.~\ref{fig-Phase_Diagram}. 
The calculations are based on a square grid $1600\,\mbox{nm}\times 1600\,\mbox{nm}$ with grid spacing $h=0.15\xi(p,T)$, where $\xi(p,T)$ is the GL coherence length, which is $\xi\approx 110\,\mbox{nm}$ for $p=15\,\mbox{bar}$ and $T/T_c=0.95$. The Nafen impurity model parameters for Nafen-90 were used as described in the previous section, and the self-consistent order parameter is scaled in units of the pure polar phase amplitude, $\Delta_{P}=\sqrt{|\alpha(T)|/2\beta_{12345}}$. 
Stationary states of the GL functional are obtained by numerical solution of the Euler-Lagrange equations for the order parameter, $A_{\alpha i}(\vr)$, using the L-BFGS numerical optimization algorithm described in Refs.~{\cite{nocedal06,reg20}}.
Convergence to a particular stationary solution generally depends on the initialization of the order parameter and asymptotic boundary conditions. To obtain the stable HQV pair we initialize with a starting HQV order parameter profile. For the initial profile we use $\Delta(\vr)=\Delta_{P}\tanh(|\vr|/\sqrt{2}\xi)$ for each HQV. Similarly, we obtain stationary SQVs by setting $x_0=0$.

In both regions of the phase diagram we find stable vortex phases that are pairs of HQVs with separation of $x_0\simeq 1030\,\mbox{nm}$ within a unit cell of dimension $d_v\approx 70\,\mu\mbox{m}$.
Figure~\ref{fig-HQV_Pair-Polar} shows the amplitudes and phases of the pair of HQVs, where $C_{\pm 0}=\nicefrac{1}{2}\Delta_{\pm}e^{i\phi_{\pm 0}}$ are complex amplitudes expressed in terms of the spin and orbital angular momentum tensors, $\lambda^{\mu\nu}_{\alpha i}$, $\mu,\nu\in\{-1,0,+1\}$~\cite{reg20}. The $2\pi$ phase winding of $\phi_{\pm 0}$ for each HQV is correlated with the zero of the corresponding amplitude.
Note that there is no difference in the vortex structure, or relative stability, for axially aligned fields $\vH=0-370G \ \hat\vOmega$, as expected for $\hat\vd\perp\vH$.

\medskip\noindent{\it Supercurrents and Spin-Polarization --}
Key signatures of the pair of HQVs in the polar phase are the broken axial symmetry of the supercurrents and the local magnetic moment associated with the two spin-polarized components of the HQV pair. 
The much denser array of line impurities leads to suppression of the polar amplitude, $\Delta^{\text{P}}(\vr)$, relative to the bulk amplitude, $\Delta_{P}$, while inhomogeneity of the order parameter induced by local pair-breaking from line impurities 
leads to weak pinning of the HQV cores at impurity sites.
The superfluid mass current in the rest frame of the excitations, i.e. the frame co-rotating with the cell, is
\begin{eqnarray} \label{eq-MassCurrents}
j_i = j_0\,\Im \left(  A^*_{\alpha j}\nabla_j A_{\alpha i} + A^*_{\alpha j}\nabla_i A_{\alpha j} + A^{*}_{\alpha i} \nabla_j A_{\alpha j} \right)
\,,
\end{eqnarray}
where $j_0 = 4m_3\,K_1/\hbar$ with the gradient coefficient given by $K_1=(7\zeta(3)/60)N_f\xi_0^2$ and $m_3$ is the atomic mass of \He. The resulting vortex supercurrents for the pair of HQVs in the polar phase break local axial symmetry, compared to the currents of an SQV, as shown in the bottom panel of Fig.~\ref{fig-HQV_Pair-Polar}.

Each member of the HQV pair is also spin polarized. The resulting zero-field magnetization density for the HQV pair is given by $\vm(\vr)=m_0(|C_{+0}|^2-|C_{-0}|^2)\hat\vOmega$ where $m_0=g'_z|\Delta_{P}|^2$. The GL material parameter $g_z'$ leads to a magnetization density of order $m_0\approx n(\gamma\hbar)\ln(E_f/k_BT_c)(\Delta_P/E_f)^2$~\cite{sau80,sau82a}. The magnetization density of the HQV pair in the ESP basis is shown in the third panel of Fig.~\ref{fig-HQV_Pair-Polar}. 
Note, that the $m^{\uparrow\uparrow}$ and $m^{\downarrow\downarrow}$ contributions to the magnetic moment exactly cancel, i.e. the net magnetic moment induced by vortex flow vanishes identically, i.e. there is no vortex-induced Barnett effect for the HQV pair in the polar phase. This result contrasts with the vortex-induced Barnett effect for vortices in \Heb\ or $^3$P$_2$ vortices in neutron matter~\cite{sau80,sau82a,hak83,reg20}. Our interpretation is that for the polar state the intrinsic orbital angular momentum of the  pair condensate vanishes~\footnote{c.f. Eq.~32 for $\cL_i$ in Ref.\onlinecite{reg20} and evaluate for any polar state.}, and thus there is no transfer of intrinsic orbital angular momentum into spin angular momentum induced by rotation.

\medskip\noindent{\it HQVs in the Chiral Phase--}
At temperatures $T< T_{c_2}$ we also find stable HQV pairs within the polar-distorted chiral phase. This phase is obtained by nucleating in-plane Cooper pairs in orbital, say $\hat{\vp}_y$, in the presence of the polar phase. Strong-coupling corrections to weak-coupling BCS theory favors a chiral A-like phase of the form $A^{\text{C}}_{\alpha i}=\DeltaC\,\hat{x}_{\alpha}\,\left(\hat{z}_{i} \pm i \epsilon\hat{y}_{i}\right)$, with the chiral axis \emph{in-plane}, e.g. in this case $\hat\vl=\pm\hat\vx$. This is also an ESP state that can host HQVs. Indeed HQVs were originally proposed as topologically stable line defects in the A-phase of bulk \He~\cite{vol76}. Consideration of the cost in dipole energy, which prefers $\hat\vd||\pm\vl$ in pure \Hea, favors collapse of the HQV pair into an SQV; however, the dipolar energy is not de-stabilizing for HQV pairs in the polar phase~\cite{min14,min16}. Similarly, for the chiral phase in Nafen, the polar component is dominant, and thus the dipolar energy does not lead to instability of the HQV pairs.

The basic structure of HQV pairs in the polar-distorted chiral phase is a simple generalization of Eq.~\ref{eq-PolarHQV_IC} obtained by the replacement, $\hat{z}_i\rightarrow{\hat{z}}_i+i\epsilon(\vr)\hat{y}_i$, 
where $\epsilon(\vr)$ defines the local chiral amplitude. Initialization of the HQV pair proceeds similarly to that for HQVs in the polar phase with the additional initialization of $\epsilon\ll 1$ to that for the equilibrium chiral phase in the absence of rotation. In addition, for $A_{\alpha i}$ with $i=x,y$, we impose a condition of zero gradient on the computational boundary.

\begin{figure}[t]
\centering
\includegraphics[width=\columnwidth]{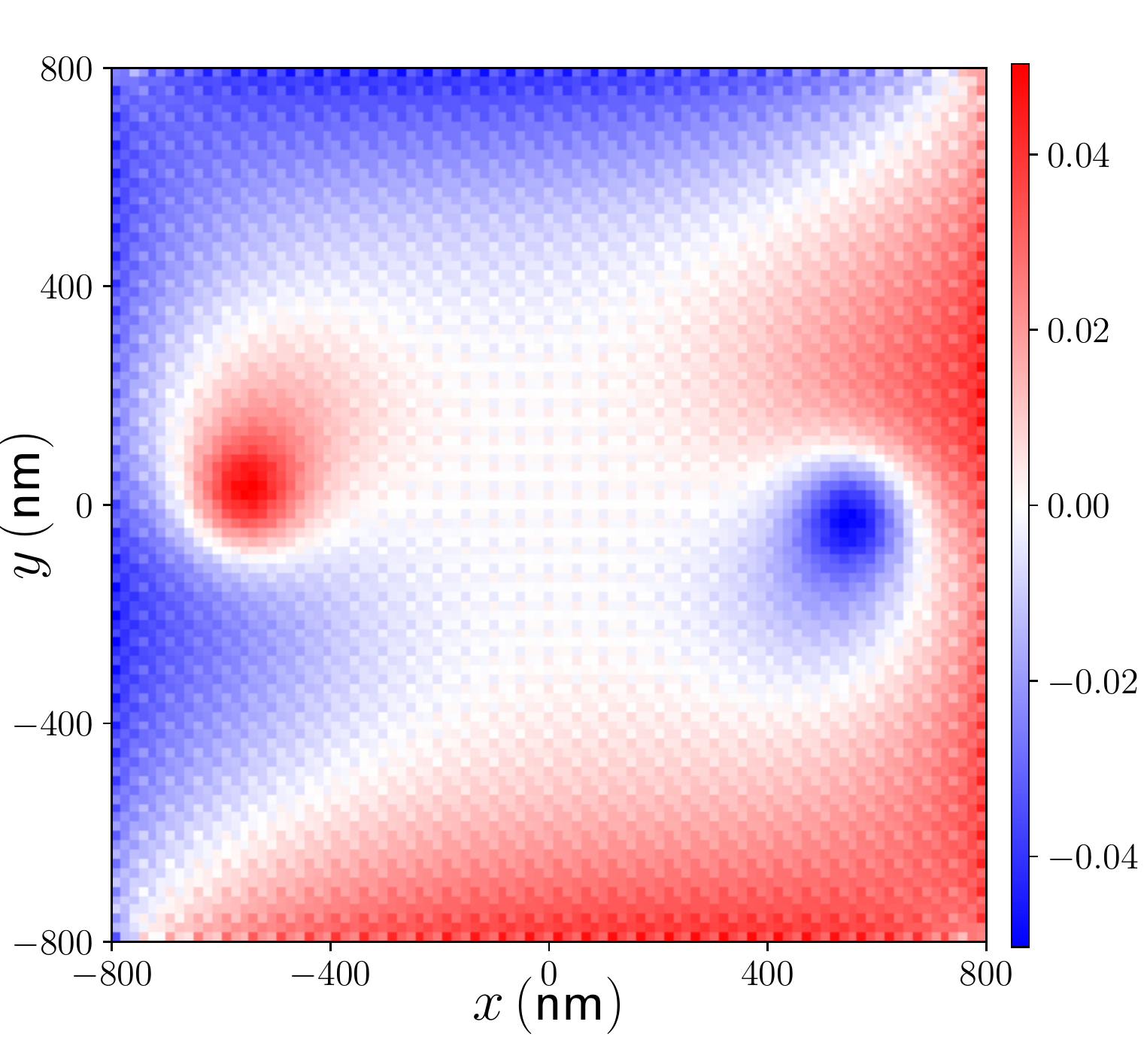}
\caption{Spontaneous supercurrent flowing along the polar axis ${\bf \hat z}$, in units of $j_0$, despite having zero phase gradient along this axis.  Taken at $p=15$bar, $T=0.85T_c$.}
\label{fig-Axial_Current-Chiral_Phase}
\end{figure}

\medskip\noindent{\it Axial Supercurrent--}
A signature of the polar-distorted chiral phase is the existence of supercurrents flowing \emph{parallel} to the vortex axis. The existence of the axial supercurrent is easily deduced from the first and third terms of Eq.~\ref{eq-MassCurrents} for the polar-distorted chiral phase with in-plane orbital components. There are two counter-propagating axial currents centered on the two HQV cores as shown in Fig.~\ref{fig-Axial_Current-Chiral_Phase}.
Similar axial currents were found earlier for the \emph{in-plane} chiral phase confined in a cylindrical pore~\cite{wim15}. However, in that case the currents reside on disclination lines pinned to the boundary wall. These currents are related to the topology of chiral phases. Any defect or boundary that suppresses the chiral phase generates edge currents confined near the defect or boundary~\footnote{c.f. Refs.~\cite{sau11,she16} and references therein.}. This fact also explains the \emph{background axial current} shown in Fig.~\ref{fig-Axial_Current-Chiral_Phase}. These are edge currents generated by the array of line impurities representing the Nafen strands~\footnote{We verified this by calculating the axial currents near the line impurities for the polar-distorted chiral phase in the absence of rotation.}.

An idea for observing the axial currents, previously proposed for the D-core vortex in \Heb~\cite{reg20}, is to inject electrons perpendicular to the nematic axis of \Hen\ from the outer cell boundary. In the presence of axial currents the electrons will be transported in \He\ parallel to the nematic axis, and can be captured and detected by imaging on the top and bottom surfaces of the cell. Such a detection of axial supercurrents in the polar-distorted chiral phase of \Hen\ would provide strong evidence of broken time-reversal symmetry and non-trivial topology in this novel phase of \He.

\medskip\noindent{\it Summary--}
The theoretical pressure-temperature phase diagram based on strong-coupling GL theory and the line-impurity pair-breaking free energy is in excellent agreement with experimental results for the normal to polar transition and a second transition identified as a polar to polar-distorted chiral phase by Dmitriev et al.~\cite{dmi15}. Both phases are ESP states that support topologically stable half-quantum vortices.
Theoretical calculations of stable HQV pair vortex arrays support the observation based on NMR of the detection of HQVs in rotating superfluid \Hen, in both the polar and polar-distorted chiral phase.
We also report signatures of the HQV pair arrays in \Hen, including anisotropic supercurrents, locally ferromagnetic HQV cores and axial supercurrents in the chiral phase whose observation would provide key signatures of HQVs in rotating superfluid \He, and confirmation of the identification of HQVs in these novel superfluid phases of \He\ confined in nematic aerogels.


\medskip\noindent{\it Acknowledgements}
We thank Wei-Ting Lin for discussions on methods for finding stationary solutions of the GL functional.
This research was supported by the National Science Foundation (Grant DMR-1508730). 

\section{Appendix: Ginzburg-Landau Theory}\label{appendix-GL_Theory}

The Ginzburg-Landau free energy functional (Eq.~\ref{eq-GL_functional}) is defined in terms of bulk, gradient, Zeeman and impurity free energy densities. The bulk condensation energy density is given by six linearly independent invariants, 
\begin{align} 
f_\mathrm{b}[A] 
&=  
\alpha(T) \Tr{A A^{\dagger}}
+\beta_{1} \left|\Tr{A A^{T}}\right|^{2}
\nonumber \\
&
+\beta_{2} \left[\Tr{A A^{\dagger}}\right]^{2}
+\beta_{3}\, \Tr{AA^{T}(AA^{T})^{*}}
\nonumber \\
&
+\beta_{4}\,\Tr{(AA^{\dagger})^{2}}
+\beta_{5}\,\Tr{AA^{\dagger}(AA^{\dagger})^{*}}
\,,
\label{eq-condensation_energy}
\end{align}
where $A^{\dag}$ ($A^{T}$) is the adjoint (transpose) of $A$.
Spatial variations of the order parameter, representing kinetic and deformation energies, are described by three linearly independent gradient terms,
\begin{equation}\label{eq-gradient_energy}
f_{\mbox{\tiny$\grad$}}[A] =
K_{1} A_{\alpha j,k}^{*} A_{\alpha j,k} +
K_{2} A_{\alpha j,j}^{*} A_{\alpha k,k} +
K_{3} A_{\alpha j,k}^{*} A_{\alpha k,j}
\,,
\end{equation}
where $A_{\alpha i,j}\equiv\grad_j A_{\alpha i}$. The gradient energy contributes to the total energy cost from impurity pair breaking, as well as kinetic and core deformation energies of quantized vortices.

In the weak coupling limit the GL material parameters, $\alpha$, $\beta_i$ and $K_j$ are given by 
\be
\alpha(p,T)=\frac{1}{3}N_f\left(T/T_c-1\right)
\,,
\ee
\ber
&&
2\beta^{wc}_1= -\beta^{wc}_2=-\beta^{wc}_3=-\beta^{wc}_4=\beta^{wc}_5
\,,
\label{eq-beta_wc}
\\
&&
\mbox{with}\quad\beta_1^{wc}=-\frac{7N_f\zeta(3)}{240 (\pi k_BT_c)^2}
\,,
\label{eq-beta1_wc}
\\
&&
\mbox{and}\quad K_1=K_2=K_3 = \frac{7 \zeta(3)}{60}N_f\,\xi^2_0
\,,
\eer
where $\xi_0 =\hbar v_f/ 2\pi \kb T_c$ is the zero-temperature Cooper pair correlation length, $N_f=m^*k_f/2\pi^2\hbar^2$ is the single-spin normal-state density of states at the Fermi level, with $p_f=\hbar k_f$ the Fermi momentum and $m^*=p_f/v_f$ is the effective mass for quasiparticles with Fermi velocity $v_f$.
The temperature-dependent GL correlation length, which is the relevant healing length scale for impurity pair breaking and vortex-core size, is given by 
\be
\xi(p,T)=\frac{\xi_{GL}}{\sqrt{1-T/T_c}}
\,,
\ee
where $\xi_{GL}=(7\zeta(3)/20)^{1/2}\xi_0$.
Strong-coupling corrections enter into the free energy functional through the fourth-order $\beta$ parameters, 
\begin{equation}
\beta_i(p,T)=\beta^{wc}_i(p,T_c)+\frac{T}{T_c}\Delta\beta^{sc}_i(p)
\,,
\label{eq-beta-parameters}
\end{equation}
where $T_c(p)$ is the bulk superfluid transition temperature, and the $\Delta\beta_i$ values were calculated in Ref.~\cite{wim19}. Note that the strong-coupling corrections are temperature and pressure dependent, which extends the standard GL theory to lower temperatures away from $T_c(p)$, and in particular can account for the A-B transition and triple point in the pressure-temperature phase diagram of bulk \He~\cite{wim16}. The most accurate theoretical results for the strong-coupling beta parameters from Ref.~\cite{wim19} are tabulated in Tables I and II of Ref.~\cite{reg20} for easy access. 

The nuclear Zeeman energy for spin-triplet pairs also plays a role in the structure of HQVs. There are two field-dependent 
contributions to the GL functional,
\begin{eqnarray}
f_\mathrm{Z_1}[A]&=&g_{z}'\Im\left\{\varepsilon_{\alpha\beta\gamma}\,\left(A\,A^{\dag}\right)_{\alpha\beta}\,H_{\gamma}\right\}
\,,
\label{eq-Zeeman_energy_linear}
\\
f_\mathrm{Z_2}[A]&=&g_{z}\,H_{\alpha}\,\left(AA^{\dag}\right)_{\alpha\beta}\,H_{\beta}
\,.
\label{eq-Zeeman_energy_quadratic}
\end{eqnarray}
where the quadratic Zeeman coupling is given by 
\be
g_z=\frac{7\zeta(3)}{48\pi^2}\frac{N_f(\gamma \hbar)^2}{\left[(1+F^a_0)\kb T_c\right]^2} > 0
\,,
\ee
with $\gamma$ being the nuclear gyromagnetic ratio of \He, and $F^a_0\approx -0.75$ the ferromagnetic exchange interaction between \He\ quasiparticles. For both the polar and chiral ESP states the quadratic Zeeman energy reduces to $f_{\mathrm{Z_2}}=g_z\Delta^2\,(\hat\vd\cdot\vH)^2$, and takes its minimum value for $\vH||\hat\vz$ and $\hat\vd\perp\hat\vz$. This is the $\hat\vd$ vector configuration favored by the dipolar energy for these two states, and is also the configuration that supports topologically stable HQVs in these two phases of \Hen.
The linear Zeeman energy has a smaller material coefficient, $g_z'=\lambda\,(\kb T_c/E_f)\,g_z \ll g_z$, where $\lambda\sim\cO(1)$, but is important in that for non-unitary states, such as the $\beta$-phase which develops in the cores of HQVs in the two ESP condensates, Eq.~\ref{eq-Zeeman_energy_linear}, is the Zeeman energy associated with the intrinsic magnetization of the non-unitary triplet state, i.e.
\be
m_\gamma = -\pder{f_{\mathrm{Z_1}}}{H_{\gamma}} = g_z'\,\Im\,\varepsilon_{\alpha\beta\gamma}\,\left(A\,A^{\dag}\right)_{\alpha\beta}
\,.
\ee

{\it Stability Analyses --}
To determine inhomogeneous equilibrium phases in the presence of impurity disorder and under rotation we minimize the strong-coupling GL functional by solving the Euler-Lagrange equations with appropriate boundary conditions, then select the lowest energy state amoung the stationary solutions, e.g. an array of HQV pairs or the SQV array. The computational procedures for \He\ in confined geometries, and under rotation via an array of topologically stable quantized vortices, are described in detail in Refs.~\cite{wim15,wim16,wim18,wim19,reg20}.
The new physics introduced here is the term in the GL free energy functional for \Hen\ representing the pair-breaking energy from the array of nematically aligned impurities, i.e. Eqs.~\ref{eq-GL_impurity}-\ref{eq-Nafen_impurity_cross-sections}.
To leading order in the order parameter, the effect of the array of line impurities embedded in superfluid \He\ is given by the quadratic terms,
\ber\label{eq-GL_functional_quadratic}
F^{(2)}_{GL} &=& \int_{V}d^3r\,
\Big\{
\alpha(p,T)\Tr{AA^{\dag}} + A_{\alpha i}\,\mrfI_{ij}(\vr)\,A^*_{\alpha j}
\\ 
             &+& K_1\,\left(
                    \grad_i A_{\alpha j}\,\grad_i A^*_{\alpha j} +
                    \grad_i A_{\alpha i}\,\grad_j A^*_{\alpha j} +
                    \grad_i A_{\alpha j}\,\grad_j A^*_{\alpha i}
                    \right)
\Big\}
\,.
\nonumber
\eer
The second-order phase transition from the normal state to the inhomogeneous superfluid state is obtained from linear instability analysis applied to Eq.~\ref{eq-GL_functional_quadratic}.
The corresponding stationarity condition, $\delta F^{(2)}_{\text{GL}}/\delta A_{\alpha i}^* = 0$, becomes
\ber\label{eq-Linearized_GL_equation}
\alpha(T)\,A_{\alpha i}(\vr) 
&+& 
A_{\alpha j}(\vr)\,\mrfI_{ji}(\vr) 
\\
&-&
K_1\,\grad^2\,A_{\alpha i}(\vr)
-
2K_{1}\,\grad_i\grad_j\,A_{\alpha j}(\vr) = 0
\nonumber
\,,
\eer
is an eigenvalue equation for the order parameter and transition temperature to the inhomogeneous superfluid phase of \Hen. For \He\ embedded in a periodic array of line impurities we impose periodic boundary conditions in the $(x,y)$ plane, and seek a solution that is translationally invariant along $z$. This allows us to express $A_{\alpha i}(\vr)$ in terms of a Fourier series, 
\be
A_{\alpha i}(\vr) = \sum_{\vQ}\,\tilde{A}_{\alpha i}(\vQ)\,e^{i\vQ\cdot\vr}
\,,
\ee
where $\vQ=\nicefrac{2\pi}{L_s}(n_x\ve_x + n_y\ve_y)$ with $n_x,n_y\in \mathbb{Z}$.
Fourier transforming Eq.~\ref{eq-Linearized_GL_equation} decouples equations for the orbital components aligned with, and orthogonal to, the nematic axis $\hat\vz$. Furthermore, for $0\le\sigma_{\parallel} < \sigma_{\perp}$ the onset of superfluidity is to the polar phase with
\be\label{eq-Polar_Instability}
A_{\alpha z}(\vr) = -\nicefrac{1}{3}N_f\,\bar{n}_s\,\xi_0\sigma_{\parallel}
                     \sum_{\vQ}\,e^{i\vQ\cdot\vr}\,
		     \frac{F(\vQ)}{\alpha(T) + K_1\,Q^2}\,A_{\alpha z}(\mathbf{0})
\,,
\ee
where $A_{\alpha z}(\mathbf{0})$ is the amplitude of the polar order parameter at the position of the impurity
centered in the unit cell. 
We include the form factor $F(\vQ)$ for the short distance structure of the line impurity on the scale of the strand diameter, $r_s\ll L_s,\xi_0$. 
The form factor regulates the short-wavelength divergence that is an artifact of the delta function representation for the local areal density. 
Specifically we model the local strand density of a single impurity by a Gaussian areal density distribution, $n_{\mbox{\tiny imp}}(\vr)=\cA\,e^{-r^2/2r_s^2}$, where $\cA=1/2\pi\sqrt{\pi}\,r_s^2$. Thus, $n_{\mbox{\tiny imp}}(\vr)\xrightarrow[r_s\rightarrow 0]{}\delta^{(2)}(\vr)$. However, in Nafen the strand dimension is small but finite, $r_s\approx 4\,\mbox{nm}\ll L_s,\xi_0$, and thus provides an ultra-violet cutoff to the Fourier sum via the structure factor,
\be
F(\vQ)\equiv\int\,d^2r\,e^{-i\vQ\cdot\vr}\,n_{\mbox{\tiny imp}}(\vr)=e^{-|\vQ|^2\,r_s^2/2}
\,.
\ee

The equation for the eigenfunction and eigenvalue (transition temperature) for the normal to polar transition is then given by Eq.~\ref{eq-Polar_Instability}
\be
\left\{
1+\nicefrac{1}{3}N_f\,\bar{n}_s\,\xi_0\sigma_{\parallel}
\sum_{\vQ}\,
	\frac{F(\vQ)}{\alpha(T_{c_1}) + K_1\,Q^2}
\right\}
A_{\alpha z}(\mathbf{0}) = 0 
\,.
\ee
For a non-trivial solution, i.e. $A_{\alpha z}(\mathbf{0})\ne 0$, we obtain the transcendental equation for $T_{c_1}$,
\be
t_{c_1} 
+
(\bar{n}_s\,\xi_0^2)\,\frac{\sigma_{\parallel}}{\xi_0}
\left[
1 + 2\sum_{n_x=1}^{\infty}\sum_{n_y=1}^{\infty}\frac{t_{c_1}\,\displaystyle{e^{-Q^2 r_s^2/2}}}{t_{c_1} + \nicefrac{7\zeta(3)}{20}\,Q^2\xi_0^2}
\right]
= 0 
\,,
\label{eq-Eigenvalue_tc1}
\ee
where $t_{c_1}\equiv T_{c_1}/T_c - 1$. The first term inside the square brackets is the contribution from the $\vQ=\mathbf{0}$ mode. The contributions from the finite wavelength modes are calculated with the leading term from the Euler-Maclaurin formula. Then $Q_1=2\pi/L_s$ provides the long-wavelength cutoff to the integral approximation to the sum, 
\ber
S
&\equiv& 
2\sum_{n_x=1}^{\infty}\sum_{n_y=1}^{\infty}\frac{t_{c_1}\,\displaystyle{e^{-Q^2 r_s^2/2}}}{t_{c_1} + \nicefrac{7\zeta(3)}{20}\,Q^2\xi_0^2}
\nonumber\\
&\simeq&
\frac{10}{7\zeta(3)\pi\bar{n}_s\xi_0^2}\times\int_{Q_1}^{\infty}\,dQ\,\frac{Q\,e^{-Q^2 r_s^2/2}}{Q^2 - Q_c^2}
\,,
\label{eq-Sum-Integral}
\eer
where $Q_c^2\equiv\nicefrac{20}{7\zeta(3)}|t_{c_1}|/\xi_0^2$. This long-wavelength scale is set by the GL coherence length at $T_{c_1}$. In particular, $Q_c/Q_1=L_s/2\pi\xi_{GL}(T_{c_1})\approx 0.08$ for $p\approx 4\,\mbox{bar}$ and $|t_{c_1}|\approx 0.1$. Thus, we have this hierarchy of wavelengths, $Q_c\ll Q_1\ll\nicefrac{\pi}{r_s}$, and thus we can drop $Q_c$ in the denominator of the integral in Eq.~\ref{eq-Sum-Integral}, in which case we obtain,
\be\label{eq-Q_Integral}
I = \int_{Q_1}^{\infty}\,\frac{dQ}{Q}\,e^{-Q^2 r_s^2/2} = \nicefrac{1}{2}\,\int_{\sqrt{\pi r_s/L_s}}^{\infty}\,\frac{dt}{t}\,e^{-t}
\,,
\ee
which is related to the Exponential integral, 
\ber
2 I(x)
&=& 
\int_{x}^{\infty}\,\frac{dt}{t}\,e^{-t}\equiv -E_i(-x)
\\
&=& -\gamma_{\text{E}} - \ln(x) - 
\int^{x}_{0}\,\frac{dt}{t}\,\left(e^{-t} - 1\right)
\,,
\eer
where $\gamma_{\text{E}}\approx 0.57722$ is the Euler-Mascheroni constant~\footnote{c.f. Sec.~8.211 and 8.212 of Ref.~\onlinecite{gradshteyn80}.}.
Thus, in the limit $0<x\ll 1$ we have $I(x)\simeq -\nicefrac{1}{2}\ln(e^{\gamma_{\text{E}}}\,x) + \cO(x)$. Noting that $L_s/r_s\gg 1$, then applying the leading order result to Eq.~\ref{eq-Q_Integral} gives us the sum over modes,
\ber
S\approx\frac{1}{2\pi}\,\frac{5}{7\zeta(3)}\,\left(\bar{n}_s\xi_0^2\right)^{-1}
\ln\left(\displaystyle{\nicefrac{e^{-2\gamma_{\text{E}}}L_s}{\pi r_s}}\right)
\,.
\eer
The resulting equation for $T_{c_1}$ then becomes,
\ber
T_{c_1} = T_c 
\left[
\frac{1-\left(\bar{n}_s\xi_0^2\right)
\frac{\sigma_{\parallel}}{\xi_0}}
     {1+\frac{5}{7\zeta(3)\pi}\,
                 \ln\left(\frac{e^{-2\gamma_{\text{E}}}}{\pi}\,\frac{L_s}{r_s}\right)
\frac{\sigma_{\parallel}}{\xi_0}
}
\right]
\,.
\eer
To leading order in $\sigma_{\parallel}/\xi_0$ we obtain Eq.~\ref{eq-Tc1} with 
\be
\beta=\frac{5}{7\zeta(3)\pi}\,
      \ln\left(\frac{e^{-2\gamma_{\text{E}}}}{\pi}\,\frac{L_s}{r_s}\right)
      \simeq 0.19\ln\left(0.10\,\frac{L_s}{r_s}\right)
\,,
\ee
which gives near perfect agreement with the experimentally measured transition temperature over the full pressure range with a pressure independent scattering cross section, $\sigma_{\parallel}=2.565\,\mbox{nm}$. 
Note that while the $\vQ\ne\mathbf{0}$ modes contribute a measurable correction to $T_{c_1}/T_c$, the dominant contribution comes from the $\vQ=\mathbf{0}$.

\medskip\noindent{\it Polar to Chiral Transition --}
The second-order polar to polar-distorted chiral transition can also be analyzed using linear stability analysis, in this case by expanding the full GL free energy functional about the polar state to leading order in the in-plane order parameter 
at temperature near $T\rightarrow T_{c_2}$, i.e. write 
\be
A_{\alpha i}(\vr)
=
\Delta^{\text{P}}(\vr)\,\hat{d}_{\alpha}\hat{z}_i
\pm i\Delta^{\text{$\perp$}}(\vr)\,\hat{d}_{\alpha}\hat{x}_i
\,, 
\ee
and expand the full $F_{GL}$ functional to quadratic order in $\Delta^{\text{$\perp$}}$. The expansion of Eq.~\ref{eq-condensation_energy} for the fourth-order terms simplifies considerably for these two ESP order parameters that have orthogonal orbital components
\ber
f_{GL} &=& \nicefrac{1}{2}\alpha_{\parallel}|\Delta^{\text{P}}(\vr)|^2 
\nonumber\\
&+& 
\left[
\alpha_{\perp} - \alpha_{\parallel}\,\frac{\beta_{245}-\beta_{13}}
                                          {\beta_{245}+\beta_{13}}
\right]
\,|\Delta^{\text{$\perp$}}(\vr)|^2
\,,
\eer 
where $\alpha_{||,\perp}=\alpha(p,T) + \nicefrac{1}{3}N_f\,\bar{n}_s\,\xi_0\,\sigma_{||,\perp}$. The first term is the condensation energy of the polar state, while the second line are the terms quadratic in the transverse component of the polar-distorted chiral phase order parameter.
Since $\Delta^{\text{P}}(\vr)$ is also periodic within the lattice model for Nafen we can express the linear instability equation for $\Delta^{\text{$\perp$}}$ in terms of Fourier mode amplitudes. In addition to the form factor for the impurity strand, $F(\vQ)$, which provides an effective cutoff at $Q\approx\pi/r_s$, the polar order parameter varies on the longer wavelength scales, i.e. $Q\approx \pi/\xi$. If we retain only the $Q=0$ contribution to the instability equation we obtain the following analytic form for the polar to chiral transition temperature, 
\be
T_{c_2}\ns=\ns T_c
\left\{1\ns-\ns\bar{n}_s\xi_0\,
\left[
\frac{
\beta_{13}
(\sigma_{\perp}\ns+\sigma_{\parallel})
\ns+\ns
\beta_{245}
(\sigma_{\perp}\ns-\sigma_{\parallel})
}
{2\beta_{13}}
\right]
\right\}
\,,
\label{eq-Tc2}
\ee
where $\beta_{ij\ldots k}=\beta_i+\beta_j+\ldots+\beta_k$. Note that $\beta_{245}$ is the combination of $\beta$ parameters that defines the condensation energy of the bulk (chiral) A-phase, while $\beta_{13}=\beta_{P}-\beta_{A}$ where $\beta_{P}=\beta_{12345}$ is the combination of $\beta$ parameters that defines the condensation energy of the bulk polar phase.
Note also that unlike Eq.~\ref{eq-Tc1}, Eq.~\ref{eq-Tc2} is a transcendental equation for $T_{c_2}$ since $\beta_{245}$ and $\beta_{13}$ are functions of pressure and temperature as defined by Eq.~\ref{eq-beta-parameters}.
The normal to polar transition, $T_{c_1}$ given by Eq.~\ref{eq-Tc1}, which includes the $Q\ne 0$ correction associated with impurity pair-breaking, describes the experimental data over the full pressure range with a pressure independent cross-section, $\sigma_{||}=2.565\,\mbox{nm}$.
The exact result for $T_{c_2}(p)$, obtained by numerical solution of the GL equations shown in Fig.~\ref{fig-Phase_Diagram}, includes the effect of the $Q\ne 0$ modes from pair-breaking by the array on line impurities. The phase boundary $T_{c_2}(p)$ shown as the solid blue line, corresponds to $\sigma_{\perp}=10.16\,\mbox{nm}$.
The $Q\ne 0$ modes from the fully established spatial variations of the polar order parameter play a more significant role in determining the polar to chiral transition line.  
For comparison, the $Q=0$ approximation for $T_{c_2}(p)$ calculated from Eq.~\ref{eq-Tc2} and Eqs.~\ref{eq-beta_wc},~\ref{eq-beta1_wc},~\ref{eq-beta-parameters} is shown in Fig.~\ref{fig-Phase_Diagram} (dashed blue line) for a pressure independent cross section $\sigma_{\perp}=8.021\,\mbox{nm}$.
%
\end{document}